\documentclass[twocolumn,useAMS]{mn2e}
\pdfoutput=1
\usepackage{natbib}

\usepackage{graphicx}
\DeclareGraphicsExtensions{.pdf,.jpg}
\usepackage{hyperref}
\usepackage{upgreek}
\usepackage[fleqn]{amsmath}
\usepackage{graphicx}

\newcommand\vx{\mbox{\boldmath $x$}}
\newcommand\vy{\mbox{\boldmath $y$}}

\newcommand\vrp{\mbox{\boldmath $r$}}
\newcommand\va{\mbox{\boldmath $\alpha$}}
\newcommand\Amat{{\textbf{\textsf{A}}}}
\newcommand\Bmat{{\textbf{\textsf{B}}}}
\newcommand\Cmat{{\textbf{\textsf{C}}}}
\newcommand\Chat{{\widetilde\Cmat}}
\newcommand\Mmat{{\textbf{\textsf{M}}}}
\newcommand\Wmat{{\textbf{\textsf{W}}}}
\newcommand\Gammamat{{\bf\Gamma}}
\newcommand\Gammahat{{\tilde\Gammamat}}
\newcommand\Imat{{\textbf{\textsf{I}}}}

\newcommand\Fmat{{\textbf{\textsf{F}}}}
\newcommand\Gmat{{\textbf{\textsf{G}}}}

\newcommand\mumat{{\bf \mu}}
\newcommand\sF{{\mathcal F}}

\newcommand\Sigcri{\Sigma_{{\rm cr},i}}

\newcommand\zeromat{{\textbf{\textsf{0}}}}
\newcommand\Transpose{{\mathrm{T}}}

\newcommand\refeq[1]{equation~(\ref{eq:#1})}

\newcommand\refeqs[2]{equations~(\ref{eq:#1}) and (\ref{eq:#2})}
\newcommand\refeqset[2]{equations~(\ref{eq:#1})--(\ref{eq:#2})}

\newcommand\refsec[1]{Section~\ref{S:#1}}

\usepackage{hyperref} 
\title[A new hybrid framework for multiplane lensing]{A new hybrid framework to efficiently model lines of sight to gravitational lenses}

\author[McCully et al.]
  {Curtis McCully,$^1$\thanks{cmccully@physics.rutgers.edu}
  Charles R.\ Keeton,$^1$ Kenneth C.\ Wong,$^{2, 3}$\thanks{EACOA Fellow} Ann I.\ Zabludoff$^2$\\
  $^1$Department of Physics and Astronomy, Rutgers University, 136 Frelinghuysen Road, Piscataway, NJ 08854, USA\\
  $^2$Steward Observatory, University of Arizona, 933 North Cherry Avenue, Tucson, AZ 85721, USA\\
  $^3$Institute of Astronomy and Astrophysics, Academia Sinica (ASIAA), PO Box 23-141, Taipei 10617, Taiwan}
  
\begin{document}
\maketitle

\begin{abstract}
In strong gravitational lens systems, the light bending is usually dominated by one main galaxy, but may be affected by other mass along the line of sight (LOS). Shear and convergence can be used to approximate the contributions from less significant perturbers (e.g. those that are projected far from the lens or have a small mass), but higher order effects need to be included for objects that are closer or more massive. We develop a framework for multiplane lensing that can handle an arbitrary combination of tidal planes treated with shear and convergence and planes treated exactly (i.e., including higher order terms). This framework addresses all of the traditional lensing observables including image positions, fluxes, and time delays to facilitate lens modelling that includes the non-linear effects due to mass along the LOS. It balances accuracy (accounting for higher-order terms when necessary) with efficiency (compressing all other LOS effects into a set of matrices that can be calculated up front and cached for lens modelling). We identify a generalized multiplane mass sheet degeneracy, in which the effective shear and convergence are sums over the lensing planes with specific, redshift-dependent weighting factors.
\end{abstract}

\section{Introduction}
\label{S:intro}

In galaxy-scale strong gravitational lens systems, there is often a single galaxy that dominates the lens potential. A few systems are compound lenses having two or three lens galaxies within the Einstein radius \citep[e.g.][]{Koopmans1608,Rusin1359,Winn0134}, and many more have significant contributions from a group or cluster environment \citep[e.g.][]{ Young81, Kundic97, Fischer98, Tonry98, Tonry99, Keeton00, Kneib00, Fassnacht06,Iva}. In all of these cases, the light bending effectively occurs in a single lens plane. If there are any massive objects along the line of sight (LOS; e.g. individual galaxies, galaxy groups or clusters, or cosmic filaments), the additional lens planes may affect the light rays in ways that cannot be ignored.

A dramatic example occurs when two galaxies at different redshifts lie close enough in projection (roughly speaking, their Einstein radii need to overlap) that both act as strong lenses. This `two-screen lensing' can produce new lensing phenomena that have been studied in detail theoretically \citep{KA88,Erdl93,PW95,Moller01,Werner08,Rhie09}. The effect is rare because it requires close alignment; it has been identified in two of the few hundred known galaxy-scale lens systems \citep{Chae2114,Gavazzi0946, Sonnenfeld12}.

It is more common to have many objects projected outside the Einstein radius \citep[e.g.][]{Tonry00}, which produce an accumulation of small perturbations that couple to the main lens. To study this scenario, one common approach is to assume that each object contributes only tidal effects---shear and convergence---to the lensing potential. Neglecting higher order effects, similar to what is used for cosmic shear studies \citep[e.g.][]{Munshi08}, makes it possible to use the statistical distribution of galaxies and large-scale structure to predict lensing perturbations \citep[e.g.][]{Seljak94,Bar-Kana96,KKS97}. In strong lens modelling, the amplitude and direction of the shear are often treated as free parameters to be optimized in individual lens systems \citep[e.g.][]{KKS97}. 

This widely used approach, which only includes tidal effects, has three possible limitations. First, it may not be appropriate to omit higher order effects beyond shear when a perturber is massive and/or close to the lens. Secondly, the shear is assumed to originate in the main lens plane, neglecting non-linear effects that arise from having mass in multiple planes \citep[see][]{Jaroszynski12}. Thirdly, lens models themselves cannot constrain any external convergence because of the mass sheet degeneracy \citep{Falco85}. To avoid biases in derived and cosmological parameters, lens model results must be adjusted after the fact to account for external convergence. It is customary to use independent data such as weak lensing \citep{Nakajima09, Fadely10} or the number density of galaxies near the lens \citep{Suyu10, Suyu13, Collett13}, although \citet{Schneider13} have questioned the efficacy of this approach as it probes the density field on scales of arcminutes, much larger than the arcsecond scales relevant for strong lensing. Even when galaxies near the lens are modelled explicitly \citep[e.g.][]{Morgan2033, CSK0435, Vuissoz2033, Fadely12} and/or external convergence is included, the mass is typically assumed to be in the main lens plane, neglecting redshift effects.

One approach to study the redshift effects due to mass along the LOS is to examine mathematical aspects of strong lensing with multiple lens planes \citep{Levine93,Kayser93,Petters95a,Petters95b,PLW}. Such studies yield rigorous results but are typically limited to general issues such as bounds on the number of images, counting rules for different types of lensed images, and classifications of caustic geometry. They do not help us account for specific, observed LOS structures in models of real lens systems.

Yet another approach is to write down the multiplane lens equation \citep[e.g.][]{BN86,Kovner87,SEF,PLW} and then perform ray-tracing calculations through appropriate three-dimensional mass distributions \citep[e.g.][]{Refsdal70,SW88a,SW88b,Jaroszynski89,Jaroszynski91,Jaroszynski92,Rauch91,Lee97,Premadi98,Wambsganss98,Wambsganss05,Hilbert07,Hilbert09, Collett13, Petkova13}. The full multiplane lens equation properly captures the redshift dependences and the couplings between redshift planes, but it can be computationally impractical. There may be hundreds of objects projected close enough to a lens to affect the light rays \citep[e.g.][]{Iva,Kurtis,Wong}, making it too expensive to evaluate the enormous number of times required in careful lens modelling. 

In this paper, we present a framework for multiplane lensing that consolidates the various approaches to provide an efficient, general way to quantify LOS effects for observed lens systems (see also \citealt{Wong}; McCully et al., in preparation). Our approach balances the accuracy of the full multiplane lens equation with the efficiency of the tidal approximation. Specifically, our framework can handle an arbitrary collection of ``main'' planes (strong lenses) that are treated exactly and tidal planes that are approximated with shear and convergence (weak lenses), at any location along the LOS. After reviewing the setup (\refsec{setup}), we analyse the lens equation and magnification tensor (\refsec{leqn}) and time delays (\refsec{tdel}) in the multiplane context.  We then examine a multiplane version of the gauge symmetry known as the mass sheet degeneracy (\refsec{masssheet}).

\section{Setup}
\label{S:setup}

Our discussion of multiplane gravitational lensing follows Chapter 9 of the book by \citet[hereafter SEF]{SEF} and Section 6.4 of the book by \citet{PLW}, which in turn draw on papers by \citet{BN86} and \citet{Kovner87}. In particular, our analysis of LOS shear in \refsec{leqn-1main} is equivalent to the discussion of the generalized quadrupole lens in Section 9.3 of SEF.

\subsection{Definitions}
\label{S:defn}

Consider $N$ galaxies with redshifts $z_i$, indexed by increasing redshift so $z_1 \le z_2 \le \ldots \le z_N < z_s$. (It is fine to have more than one galaxy at a given redshift.)  The source is in plane $N+1$, which is labelled with the index $s$. Let $D_i$ and $D_{is}$ be the angular diameter distances from the observer to galaxy $i$ and from galaxy $i$ to the source (respectively). For $i<j$ let $D_{ij}$ be the angular diameter distance from galaxy $i$ to galaxy $j$.

Let galaxy $i$ have lensing potential $\phi_i(\vx_i)$ and surface mass density $\Sigma_i(\vx_i)$. The lensing effects are functions of the angular position $\vx_i$ of a light ray as it passes through plane $i$, which in general is not the same as the observed position on the sky. The position $\vx_i$ depends on how the light is bent by other planes, as characterized by the lens equation (\ref{eq:multi-x}). The lensing potential and surface mass density are related by the Poisson equation
\begin{equation}
  \nabla^2\phi_i(\vx_i) = 2 \frac{\Sigma_i(\vx_i)}{\Sigcri}\ ,
\end{equation}
where the critical surface density for lensing for plane $i$ is
\begin{equation}
  \Sigcri = \frac{c^2}{4\uppi G}\ \frac{D_s}{D_i D_{is}}\ .
\end{equation}
The deflection angle from galaxy $i$ is then
\begin{equation}
  \va_i(\vx_i) = \nabla\phi_i(\vx_i)\,.
\end{equation}
It is useful to introduce the matrix of second derivatives, or the `tidal tensor':
\begin{equation}
  \Gammamat_i = \frac{\partial\va_i}{\partial\vx_i}
  = \left[\begin{array}{cc}
    \kappa_i + \gamma_{\mathrm{c},i} & \gamma_{\mathrm{s},i} \\
    \gamma_{\mathrm{s},i} & \kappa_i - \gamma_{\mathrm{c},i}
  \end{array}\right] ,
\end{equation}
where we define the convergence ($\kappa$) and shear ($\gamma$) components from galaxy $i$:
\begin{eqnarray}
  \kappa_i &=& \frac{1}{2} \left( \frac{\partial^2\phi_i}{\partial x_i^2} +
    \frac{\partial^2\phi_i}{\partial y_i^2} \right) , \\
  \gamma_{\mathrm{c},i} &=& \frac{1}{2} \left( \frac{\partial^2\phi_i}{\partial x_i^2} -
    \frac{\partial^2\phi_i}{\partial y_i^2} \right) , \\
  \gamma_{\mathrm{s},i} &=& \frac{\partial^2\phi_i}{\partial x_i \partial y_i}\ .
\end{eqnarray}
Note that the convergence can be obtained from the trace of $\Gammamat$, while the shear components are given by the traceless, symmetric part of $\Gammamat$.

We can Taylor expand the lens potential for a perturbing galaxy about the centre of the main lens galaxy as
\begin{equation}
\phi(\vx) = \phi(0) + \alpha^a(0) x^a + \frac{1}{2}\Gamma^{ab} x^a x^b+ \frac{1}{6} \sF^{abc} x^a x^b x^c + \cdots 
\label{eq:phiser}
\end{equation}
where $a,b,c$ are vector or tensor component indices and we have adopted the Einstein notation of summing over repeated indices. $\sF$ is the flexion tensor of third derivatives defined by
\begin{equation}
\sF^{abc} \equiv  \left. \frac{\partial^3 \phi}{\partial x^a \partial x^b \partial x^c} \right|_{x=0}.
\end{equation}
In \refeq{phiser}, the $\phi(0)$ term is the zeropoint of the potential, which is unobservable.  The $\va(0)$ term corresponds to a uniform deflection that is degenerate with a translation of the source plane coordinates. Thus, the first significant term is the second-order one.  If we can neglect higher order terms and truncate the expansion at second order, we have
\begin{eqnarray} \label{eq:shearapprox}
  \phi_i (\vx_i) &\approx& \frac{1}{2} \vx_i \cdot \Gammamat_i(0) \vx_i\,, \\
  \va_i  (\vx_i) &\approx& \Gammamat_i(0) \vx_i\,, \\
  \Gammamat_i(\vx_i) &\approx& \Gammamat_i(0)\,.
\end{eqnarray}
This defines the tidal approximation, which we employ for all planes in which the higher-order terms beyond shear are sufficiently small. (We quantify the accuracy of the tidal approximation in a forthcoming paper; McCully et al. in preparation) In the remainder of the paper we drop $(0)$ for simplicity. We refer to planes that employ the tidal approximation as `tidal planes,' and planes that are treated exactly as `main planes.'

For illustration, the lensing potential of a point mass is given by
\begin{equation}
\phi(\vx) = R_{\mathrm{E}}^2 \ln \left| \vx - \vrp_{\mathrm{p}} \right| 
\end{equation}
where $\vrp_{\mathrm{p}}$ and $R_{\mathrm{E}}$ are the position and Einstein radius of the perturber, respectively. If we let $|\vx| = x$, $|\vrp_{\mathrm{p}}| = r_\mathrm{p}$, and $\theta$ be the angle between the perturber and the image position as measured from the origin, then we can rewrite the potential using the law of cosines as
\begin{equation}
\phi(x,\theta) = \frac{1}{2}R_{\mathrm{E}}^2 \ln\left(r_{\mathrm{p}}^2 + x^2 - x r_{\mathrm{p}} \cos\theta\right).
\end{equation}
If we assume the projected offset of the perturber is large compared to the image positions ($r_{\mathrm{p}} \gg x$), then we can expand the logarithm as
\begin{multline}
\phi(x,\theta) \approx R_{\mathrm{E}}^2 \left[ \ln(r_{\mathrm{p}}) - \cos(\theta) \frac{x}{r_{\mathrm{p}}}  - \frac{1}{2} \cos(2\theta) \frac{x^2}{r_{\mathrm{p}}^2} \right. \\
\left. - \frac{1}{3}\cos(3\theta)\frac{x^3}{r_{\mathrm{p}}^3}  + \cdots\right].
\end{multline}
We see that a point mass has $\Gammamat \propto R_E^2/r_p^2$ and $\sF \propto R_E^2/r_p^3$.

\subsection{Multiplane lensing}
\label{S:multiplane}

The lens equation is constructed by working ``backwards'' from the observer, through the lens planes one by one, until we reach the source.  If $\vx_j$ is the position in plane $j$, we have (see equation~9.7a of SEF, and equation~6.29 of \citealt{PLW})
\begin{equation}
  \vx_{j} = \vx_1 - \sum_{i=1}^{j-1} \beta_{ij} \va_i(\vx_i)\,,
 \label{eq:multi-x}
\end{equation}
where
\begin{equation} \label{eq:beta}
  \beta_{ij} = \frac{D_{ij} D_s}{D_j D_{is}}\ .
\end{equation}
Note that the lens equation for plane $j$ depends on all planes in front of $j$ ($i < j$), so this amounts to a recursion relation that we can use to start with angular coordinates on the observer's sky ($\vx_1$) and work our way up in redshift until we reach the source plane ($\vx_s = \vx_{N+1}$). Some authors \citep[e.g.][]{Seitz94,Hilbert09} write the recursion relation in a different form, but we find \refeq{multi-x} to be useful.

The Jacobian matrix for the mapping between the coordinates on the sky and the coordinates in plane $j$ is
\begin{equation} \label{eq:multi-A}
  \Amat_{j} \ =\ \frac{\partial\vx_{j}}{\partial\vx_1}
  \ =\ \Imat - \sum_{i=1}^{j-1} \beta_{ij} \frac{\partial\va_i}{\partial\vx_i}
       \frac{\partial\vx_i}{\partial\vx_1}
  \ =\ \Imat - \sum_{i=1}^{j-1} \beta_{ij} \Gammamat_i \Amat_i\,,
\end{equation}
where $\Imat$ is the $2\times2$ identity matrix.  The lensing magnification tensor is the inverse of the Jacobian matrix for the source plane: $\mumat = \Amat_s^{-1}$.

The general form for the multiplane time delay is (see equation 6.22 of \citealt{PLW})
\begin{equation}
  T = \sum_{i=1}^{s-1} \tau_{i\,i+1} \left [ \frac{1}{2} |\vx_{i+1} - \vx_i|^2 - \beta_{i\,i+1} \phi_i(\vx_i)\right ],
\label{eq:full_t}
\end{equation}
where 
\begin{equation}
\tau_{ij} =\frac{1+z_i}{c} \frac{D_i D_j}{D_{ij}}
\end{equation}
is a distance combination with dimensions of time.  We can omit the redshift dependence if we measure $D_i$, $D_j$, and $D_ij$ as comoving rather than angular diameter distances.

Throughout the derivation we use the following identities from the definitions of $\beta_{ij}$ and $\tau_{ij}$ (see Section 6.4.1 in \citealt{PLW}):
\begin{eqnarray}
  \beta_{is} &=& 1 \qquad(\forall i)\,, \label{eq:b_1}\\
  \tau_{is} &=& \beta_{ij} \tau_{ij} \qquad(\forall ij)\,, \label{eq:t_b} \\
  \frac{1}{\tau_{ik}} &=& \frac{1}{\tau_{ij}} + \frac{1}{\tau_{jk}} \qquad (i<j<k)\, \label{eq:t_ik}.
\end{eqnarray}
Also, to simplify the notation we define versions of $\beta$ and $\tau$ with a single subscript as
\begin{equation}
  \beta_i \equiv \beta_{i\,i+1}\,,
  \qquad
  \tau_i \equiv \tau_{i\,i+1}\,.
\end{equation}

\section{Lens Equation and Magnification Tensor}
\label{S:leqn}

In this section we work with the multiplane lens equation and magnification tensor. We start by using the tidal approximation for all planes other than the plane containing the main lens galaxy. We then generalize to arbitrary combinations of tidal and main planes.

\subsection{One `main' plane}
\label{S:leqn-1main}

\begin{figure*}
\begin{center}
\includegraphics[width=1.0\textwidth]{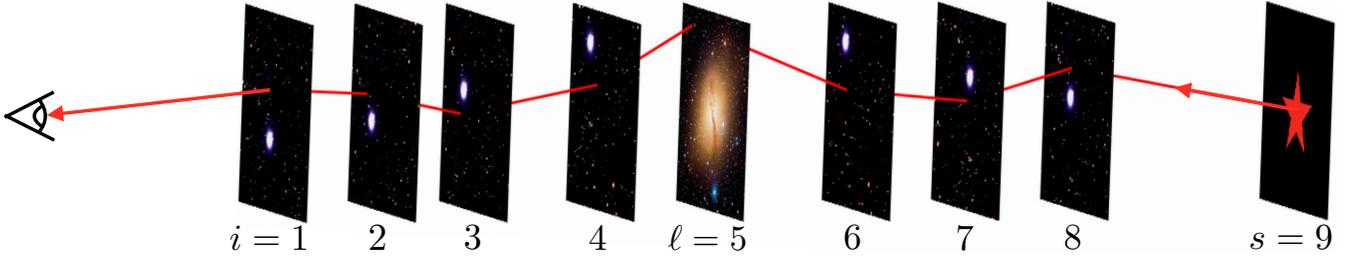}
\caption{Schematic diagram of multiplane lensing (not to scale).  The light bending is dominated by a single main plane ($\ell=5$) but affected by additional tidal planes in the foreground and background of the main plane.  Here the source is in plane $s=9$, but our framework can handle an arbitrary number of planes. Image credits: Centaurus A - CFHT/Coelum (J.-C. Cuillandre \& G. Anselmi).
}
\label{fig:1plane}
\end{center}
\end{figure*}
Suppose there is a single `main' lens plane ($i = \ell$) and all other galaxies can be treated with the tidal approximation as illustrated in Fig. \ref{fig:1plane}.  (This case has been studied previously by \citealt{Kovner87} and SEF.)  Using \refeq{shearapprox}, we can write the recursion relations for the position and Jacobian matrix as
\begin{eqnarray}
  \vx_j   &=& \vx_1 - \sum_{i=1,i\ne\ell}^{j-1} \beta_{ij} \Gammamat_i \vx_i
                    - \beta_{\ell j} \va_\ell(\vx_\ell)\,, \\
    \label{eq:vxj}
  \Amat_j &=& \Imat - \sum_{i=1,i\ne\ell}^{j-1} \beta_{ij} \Gammamat_i \Amat_i
                    - \beta_{\ell j} \Gammamat_\ell(\vx_\ell) \Amat_\ell\,.
    \label{eq:Amatj}
\end{eqnarray}
We separate the terms with $i = \ell$ and write $\va_\ell$ and $\Gammamat_\ell$ explicitly because we do \emph{not} use the tidal approximation for the main plane.

It is interesting to consider the position $\vx'_j$ and Jacobian matrix $\Bmat_j$ that we would get if we were to omit the main plane.  These quantities must be used with care because they do not include contributions from the main plane (which will be added back in later), but they will prove to be valuable.  These modified quantities have the form
\begin{eqnarray}
  \vx'_j   &=& \vx_1 - \sum_{i=1,i\ne\ell}^{j-1} \beta_{ij} \Gammamat_i \vx'_i \,,
    \label{eq:vx'j} \\
  \Bmat_j &=& \Imat - \sum_{i=1,i\ne\ell}^{j-1} \beta_{ij} \Gammamat_i \Bmat_i \,.
    \label{eq:Bmatdef}
\end{eqnarray}
In the foreground of the main lens plane ($j \le \ell$), we clearly have $\vx'_j = \vx_j$ and $\Bmat_j = \Amat_j$ because the trajectory has not yet been affected by the main plane.  (Recall that we trace a light ray backwards from the observer.)  The situation is different; however, in the background of the main lens plane ($j > \ell$).  Taking the difference between \refeqs{Amatj}{Bmatdef}, we have
\begin{equation}
  \Amat_j-\Bmat_j = - \beta_{\ell j} \Gammamat_\ell \Amat_\ell
    - \sum_{i=\ell+1}^{j-1} \beta_{ij} \Gammamat_i (\Amat_i-\Bmat_i) \,.
\end{equation}
Note that the sum now includes only terms with $i > \ell$, because $\Amat_i-\Bmat_i=0$ for $i \le \ell$.  Now if we multiply through by $(- \Gammamat_\ell \Bmat_\ell)^{-1}$ from the right and use the fact that $\Amat_\ell = \Bmat_\ell$, we obtain
\begin{eqnarray} 
  \Cmat_{\ell j} &\equiv & (\Amat_j-\Bmat_j) (- \Gammamat_\ell \Bmat_\ell)^{-1} \\
   &= & \beta_{\ell j} \Imat
    - \sum_{i=\ell+1}^{j-1} \beta_{ij} \Gammamat_i \Cmat_{\ell i} .
\label{eq:Cmatdef}
\end{eqnarray}
Equation (\ref{eq:Cmatdef}) is a recursion relation for $\Cmat_{\ell j}$ that involves only LOS effects, specifically only planes in between the main plane and plane $j$.  In other words, $\Cmat_{\ell j}$ is independent of the main lens.  There is, of course, a dependence on the main lens in converting between $\Cmat_{\ell j}$ and $\Amat_j$ with
\begin{equation}
  \Amat_j = \Bmat_j - \Cmat_{\ell j} \Gammamat_\ell \Bmat_\ell\,.
\end{equation}

The matrices $\Bmat_j$ and $\Cmat_{\ell j}$ turn out to have an additional use when we consider the positions.  Returning to \refeqs{vxj}{vx'j} and writing out terms, we find that in the tidal approximation we have the simple relation
\begin{equation}
  \vx'_j = \Bmat_j \vx_1 \,,
\end{equation}
for all $j$.  In the foreground ($j \le \ell$) we of course have $\vx_j = \vx'_j$.  In the background ($j > \ell$), the positions $\vx_j$ and $\vx'_j$ are different, and in fact we have
\begin{equation}
  \vx_j \ =\ \vx'_j - \Cmat_{\ell j} \va_\ell(\vx_\ell)
  \ =\ \Bmat_j \vx_1  - \Cmat_{\ell j} \va_\ell(\vx_\ell) \,.
  \label{eq:singlelenseq}
\end{equation}
Note that the deflection depends on the position in the main lens plane $\vx_\ell$, not the observed sky plane $\vx_1$. Therefore tidal effects from foreground planes couple to the deflection from the main lens plane and cannot be mimicked by a standard, linear shear term in the lens plane.\footnote{Equation (\ref{eq:singlelenseq}) can be made formally equivalent to the standard single-plane lens equation with a suitable transformation of the lens potential \citep{Schneider97}. In that case, however, the effective mass model differs from the true mass distribution of the lens. The fact that the lens potential must be distorted further emphasizes that tidal contributions from foreground planes create important non-linear effects.} The resulting non-linear effects are important for the multiplane mass sheet degeneracy and for lens modelling (see Sections \ref{S:masssheet} and \ref{S:conclusions}).

To summarize, in the case of a single main plane plus a collection of planes that can be treated with the tidal approximation, we can separate the full multiplane lensing analysis into pieces that depend only on the LOS ($\Bmat_\ell$, $\Bmat_s$, and $\Cmat_{\ell s}$) and pieces that depend on the main lens plane ($\va_\ell$ and $\Gammamat_\ell$, both of which are evaluated at the position $\vx_\ell = \Bmat_\ell \vx_1$).  We can combine the pieces into the lens equation and Jacobian matrix as follows:
\begin{eqnarray}
  \vx_s   &=& \Bmat_s \vx_1 - \Cmat_{\ell s} \va_\ell(\Bmat_\ell \vx_1) \,,
    \label{eq:xfin1} \\
  \Amat_s &=& \Bmat_s - \Cmat_{\ell s} \Gammamat_\ell \Bmat_\ell \,.
    \label{eq:Afin1}
\end{eqnarray}
This represents a \emph{complete} description of the multiplane lensing in this scenario; there are no approximations involved in the treatment of multiplane lensing itself.  The only approximation used here is the tidal approximation for the perturbing galaxies.

The multiplane lens equation (\ref{eq:xfin1}) is identical to the quadrupole lens equation in SEF and equivalent to the results from \citet{Kovner87} and \citet{Bar-Kana96}. From a formal standpoint, the equation can be made to look like the standard single-plane lens equation through a suitable change of variables \citep{Bar-Kana96,Schneider97,CRKanal}. From a practical standpoint, however, the change of variables is of limited use because lens modelling needs to use \emph{observed} coordinates. (Observed and scaled coordinates can be related to one another only if the transformation matrices are known, in which case one might as well use equation \ref{eq:xfin1}.)

\subsection{Small-shear limit}
\label{S:leqn-small}

It is instructive to consider the preceding analysis in the limit where all the LOS shears are small.  If we make Taylor series expansions and work to linear order in the LOS shears, we obtain
\begin{eqnarray}
  \Bmat_s    &\approx& \Imat - \Gammamat_{\rm tot}\,, \\
  \Bmat_\ell &\approx& \Imat - \Gammahat_{\rm f} \,, \\
  \Cmat_{\ell s}    &\approx& \Imat - \Gammahat_{\rm b} ,
\end{eqnarray}
where
\begin{equation} 
  \Gammamat_{\rm tot} = \sum_{i=1,i\neq\ell}^{N} \Gammamat_i
  \label{eq:fgbg1}
\end{equation}
are simple sums of the foreground and background tidal tensors (with uniform weighting), while
\begin{eqnarray} \label{eq:fgbg2}
  \Gammahat_{\rm f} = \sum_{i=1}^{\ell-1} \beta_{i \ell} \Gammamat_i
  \quad\mbox{and}\quad
  \Gammahat_{\rm b} = \sum_{i=\ell+1}^{N} \beta_{\ell i} \Gammamat_i
\end{eqnarray}
are sums where the different planes have different weighting factors $\beta_{i \ell} \ne 1$ and $\beta_{\ell i} \ne 1$. The different weighting factors between $\Gammamat$ and $\Gammahat$ will be important for the discussion of the mass sheet degeneracy (Section \ref{S:masssheet}). Note that \citet{Wong} used $\Gammamat_{\rm tot}$ to characterize environmental effects for observed lenses.  The sums above are discretized versions of the integrals used in cosmic shear calculations \citep[e.g.][]{Munshi08}.

\subsection{Multiple `main' planes}
\label{S:leqn-Nmain}

\begin{figure*}
\begin{center}
\includegraphics[width=1.0\textwidth]{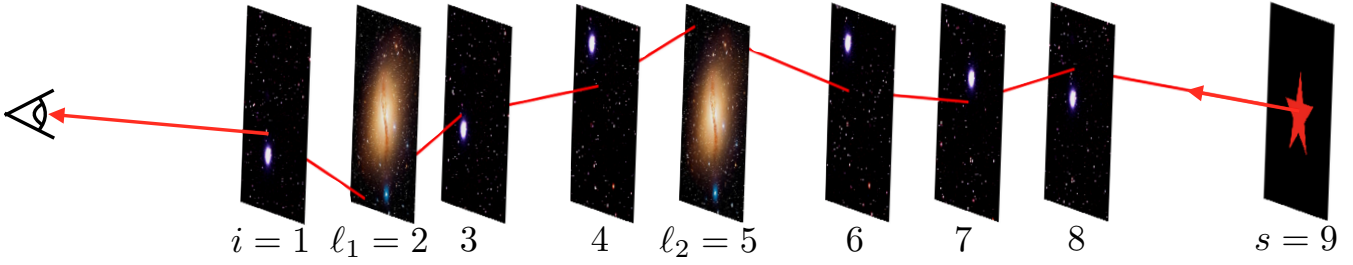}
\caption{ Similar to Fig. \ref{fig:1plane}, but showing a case with two main planes ($\ell_1 = 2$ and $\ell_2 = 5$). Image credits: Centaurus A - CFHT/Coelum (J.-C. Cuillandre \& G. Anselmi).}
\label{fig:2plane}
\end{center}
\end{figure*}

We now extend the framework to allow arbitrary combinations of main planes (which are given full treatment) and tidal planes, illustrated in Fig. \ref{fig:2plane}. We do not make any particular assumptions about how the planes are distributed in redshift; there may be 0, 1, or many tidal planes in between any two main planes. As noted above, more than one galaxy may be at a given redshift. Our notation is as follows: Roman letters $(i,j)$ are used to sequentially index all planes (both main and shear). Greek letters $(\mu,\nu)$ are used to sequentially index main planes only. Also, $\ell_\mu$ denotes the Roman index of the main plane $\mu$; in other words, $\{\ell_1,\ell_2,\ldots,\ell_\mu,\ldots\}$ are the indices of the main planes. The source plane counts as a main plane, but with index $s = N+1$. 

To set the stage, let us re-examine the multiplane lens equations for the case in which all planes are main (equations \ref{eq:multi-x} and \ref{eq:multi-A}) and the case with a single main plane (equations \ref{eq:xfin1} and \ref{eq:Afin1}).  The first term in each case represents what would happen if the main planes were not present: in \refeq{multi-x} this is characterized by the identity matrix because if we remove all planes we are left with no lensing; while in \refeq{xfin1} there is distortion from all the tidal planes, which is characterized by the matrix $\Bmat_s$.  The terms in the sums represent the combined contributions from the main plane(s) in the foreground of the plane being evaluated.  In \refeq{multi-x} the light ray experiences no distortions in between planes, so the connecting factor is just a scalar ($\beta_{ij}$) that encodes the relative distances between planes $i$ and $j$.  In \refeq{xfin1}, by contrast, the light ray may be sheared in between main planes, so the connecting factor becomes a matrix ($\Cmat_{\ell s}$) that includes not only the distance factors but also the shears in between the main planes.

We can now understand the form of the lens equations for a general combination of main and tidal planes:
\begin{eqnarray}
  \vx_{i} &=& \Bmat_i \vx_1 - \sum_{\ell \in \{\ell_\mu<i\}} \Cmat_{\ell i} \va_\ell(\vx_\ell) \,,
    \label{eq:xfinN} \\
  \Amat_{i} &=& \Bmat_i - \sum_{\ell \in \{\ell_\mu<i\}} \Cmat_{\ell i} \Gammamat_\ell \Amat_\ell \,.
    \label{eq:AfinN}
\end{eqnarray}
Again note that the deflections depend on the positions in the main planes $\vx_\ell$. Also, these sums only include main planes.  At each step in the recursion, $\va_\ell$ and $\Gammamat_\ell$ are to be evaluated at the position $\vx_\ell$. The matrix $\Bmat_\ell$ represents the net effects of the tidal planes in between the observer and the main plane with index $\ell$, which can be found recursively as follows:
\begin{equation}
\Bmat_j = \Imat - \sum\limits_{i=1,i\not \in \{\ell_\mu\}}^{j-1}\beta_{ij}\Gammamat_i \Bmat_i
\end{equation}
where this sum does not include any of the main planes (even if they happen to lie between the observer and plane $j$). The matrix $\Cmat_{\ell j}$ represents the net effects of the tidal planes in between the main plane $\ell$ and plane $j$ whose recursion relation is
\begin{equation}
\Cmat_{\ell j} = \beta_{\ell j} \Imat - \sum\limits_{i=\ell+1, i\not\in\{\ell_\mu\}}^{j-1} \beta_{i j}\Gammamat_i\Cmat_{\ell i},
\end{equation}
where again this sum only includes tidal planes. Note that $\Bmat_j$ and $\Cmat_{\ell j}$ are defined for arbitrary $j$, but \refeqs{xfinN}{AfinN} show that only the matrices associated with main planes need to be stored for later use. The benefit of this approach for lens modelling is that the bulk of the computational effort goes into determining $\Bmat_{\ell_\mu}$ and $\Cmat_{\ell_\mu \ell_\nu}$, but that step needs to be done only once. Once those matrices are stored, the mass model in the main plane(s) can be varied without having to recompute the full LOS.

\section{Time Delay}
\label{S:tdel}

We now turn to time delays.  As before, we start with a single main plane plus a collection of tidal planes, and then generalize to an arbitrary combination of tidal and main planes.  To set the context, it is useful to recall the classic expression for the time delay in single-plane lensing.  The single-plane time delay can be written in several different forms, the most familiar of which is
\begin{equation}
T \propto \frac{1}{2} |\vx-\vx_s|^2 - \phi(\vx)\,.
\end{equation}
We can expand the quadratic term as 
\begin{equation}
T \propto \frac{1}{2} ( \vx^2 - \vx \cdot \vx_s - \vx_s \cdot \vx + \vx^2_s) -  \phi(\vx)\,. \label{eq:t_quad}
\end{equation}
In terms of the deflection angle $\va$, we can rewrite this as 
\begin{equation}
T \propto \frac{1}{2} | \va|^2 - \phi(\vx)\,.
\end{equation}
We can even mix these two forms giving
\begin{equation}
T \propto \frac{1}{2} (\vx-\vx_s)\cdot \va - \phi(\vx)\,.
\label{eq:t_alpha}
\end{equation}
While these forms may look rather distinct, they are all equivalent.  We will see below how the different forms are useful.

\subsection{Single `main' plane}
\label{S:tdel-1main}
The general expression for the multiplane time delay depends explicitly on \textit{all} of the $\vx_j$ and $\phi_j$. Our goal is to write the time delay in terms of $(\vx, \vx_s, \phi_\ell)$ or equivalently $(\vx, \va_\ell,\phi_\ell)$. To that end, we substitute for the position coordinates, explicitly separate out the main plane lens potential, and implement the tidal approximation for all other planes ($\phi_j\approx \frac{1}{2} \vx_j \cdot\Gammamat_j\vx_j$). This yields
\begin{equation}
T =\sum\limits_{i=1}^{s-1} \frac{1}{2} \tau_{i} \left[ \vx_{i+1} - \vx_i \right]^2 -\tau_{\ell s} \phi_\ell(\vx_{\ell}) - \sum\limits_{i=1,i\neq\ell}^{s-1} \frac{1}{2} \tau_{i} \beta_{i} \vx_i\cdot\Gammamat_i\vx_i .
\label{eq:t1}
\end{equation}
We would like to eliminate $\Gammamat_i$, so it is necessary to digress to derive a few useful identities. We start by examining
\begin{equation}
\Bmat_{j+1}-\Bmat_j = - \beta_{j} \Gammamat_j\Bmat_j - \sum\limits_{i\neq\ell}^{j-1} (\beta_{i\,j+1}-\beta_{ij})\Gammamat_i\Bmat_i,
\end{equation}
and 
\begin{equation}
\Bmat_{j}-\Bmat_{j-1} = - \beta_{j-1} \Gammamat_{j-1}\Bmat_{j-1} - \sum\limits_{i\neq\ell}^{j-2} (\beta_{ij}-\beta_{i\,j-1})\Gammamat_i\Bmat_i.
\label{eq:a-}
\end{equation}
Combining these and using \refeq{t_b}, we can cancel the sum to obtain (see also \citealt{Seitz94})
\begin{equation}
\Bmat_{j+1} = \left[\left(1 +\frac{\tau_{j-1}}{\tau_{j}}\right)\Imat-\beta_{j} \Gammamat_j \right] \Bmat_j - \frac{\tau_{j-1}}{\tau_{j}} \Bmat_{j-1}.
\label{eq:Bjidentity}
\end{equation}
Rearranging, we can solve for $\Gammamat_j$:
\begin{equation}
\beta_{j} \Gammamat_j  = \left(1 +\frac{\tau_{j-1}}{\tau_{j}}\right)\Imat - \Bmat_{j+1} \Bmat^{-1}_{j} -  \frac{\tau_{j-1}}{\tau_{j}} \Bmat_{j-1}\Bmat^{-1}_{j}.
\label{eq:b_g}
\end{equation}
Following the same procedure yields a similar result for $\Cmat_{\ell j}$:
\begin{equation}
\beta_{j} \Gammamat_j  = \left(1 +\frac{\tau_{j-1}}{\tau_{j}}\right)\Imat - \Cmat_{\ell \,j+1} \Cmat^{-1}_{\ell j} -  \frac{\tau_{j-1}}{\tau_{j}} \Cmat_{\ell \,j-1}\Cmat^{-1}_{\ell j}.
\label{eq:c_g}
\end{equation}
We now multiply the lens equation (\ref{eq:singlelenseq}) by $\beta_{j} \Gammamat_j$ from the left, and use \refeq{b_g} when $\Gammamat_i$ multiplies $\Bmat_i$ and \refeq{c_g} when $\Gammamat$ multiplies $\Cmat_{\ell j}$ yields 
\begin{multline}
\beta_j \Gammamat_j\vx_j = \left[\left(1 + \frac{\tau_{j-1}}{\tau_j}\right) \Bmat_j - \Bmat_{j+1} - \frac{\tau_{j-1}}{\tau_j}\Bmat_{j-1} \right] \vx_1 \\
- \left[\left(1 + \frac{\tau_{j-1}}{\tau_j}\right) \Cmat_{\ell j} - \Cmat_{\ell\,j+1} - \frac{\tau_{j-1}}{\tau_j}\Cmat_{\ell\,j-1} \right]\va_\ell(\vx_\ell).
\end{multline}
Collecting terms and again using \refeq{singlelenseq} yields
\begin{equation}
\beta_j\Gammamat_j \vx_j = \left(1+\frac{\tau_{j-1}}{\tau_{j}}\right)\vx_j - \vx_{j+1} - \frac{\tau_{j-1}}{\tau_{j}} \vx_{j-1}.
\label{eq:x_g}
\end{equation}
Using this relation in \refeq{t1} gives
\begin{multline}
T = \sum\limits_{i=1}^{s-1} \frac{1}{2} \tau_{i} \left[ \vx_{i+1}^2 - 2 \vx_{i+1} \cdot \vx_{i} +  \vx_i^2  \right] -\tau_{\ell s} \phi_\ell(\vx_{\ell})
\\
- \sum\limits_{i=1,i\neq\ell}^{s-1} \frac{1}{2} \tau_{i} \vx_i \cdot\left[\left(1+\frac{\tau_{i-1}}{\tau_{i}}\right)\vx_i - \vx_{i+1} - \frac{\tau_{i-1}}{\tau_{i}} \vx_{i -1} \right] .
\label{eq:t_sum}
\end{multline}
The identity term in the second sum is identical to the second quadratic term in the first sum but with opposite sign. Also, the first $i,i+1$ cross term in the first sum matches the $i,i+1$ cross term in the second sum. These terms cancel except for the main plane term $j=\ell$ that we explicitly removed from the second sum.

The other terms in the first sum have the same form as the remaining terms in the second sum, but with indices decremented by 1. We therefore reindex the remaining terms in the second sum with $i \rightarrow i+1$. These terms become
\begin{equation}
\sum\limits^{s-1}_{i=1,i\neq\ell} \tau_{i-1} \vx_{i}^2  \rightarrow \sum\limits^{s-2}_{i=0,i\neq \ell-1} \tau_{i} \vx_{i+1}^2 
\end{equation}
and 
\begin{equation}
\sum\limits^{s-1}_{i=1,i\neq\ell} \tau_{i-1} \vx_{i} \cdot \vx_{i-1}. \rightarrow \sum\limits^{s-2}_{i=0,i\neq \ell-1} \tau_{i} \vx_{i+1} \cdot \vx_i.
\end{equation}
These match the terms in the first sum but have opposite sign and therefore all of the sums cancel. The only surviving terms are $s-1$ and $\ell -1$ terms from removing the main plane and reindexing. There is also an $i=0$ term from the second reindexed sum. This term would have $\tau_{0,1}$ as a coefficient. Taking the zero plane to be the observer, we have $D_0=0$ and therefore $\tau_{0,1}=0$. This leaves us with
\begin{multline}
T = \frac{1}{2} \left[ \tau_{\ell} \vx_\ell \cdot \left(\vx_\ell  -  \vx_{\ell +1} \right) + \tau_{\ell-1} \vx_\ell\cdot \left(\vx_\ell  - \vx_{\ell-1} \right) \right.\\
\left.+ \tau_{s-1} \vx_s \cdot \left(\vx_s  - \vx_{s-1} \right) \right] -\tau_{\ell s} \phi_\ell(\vx_{\ell}).
\label{eq:Txdiff}
\end{multline}

The terms in \refeq{Txdiff} are of the form $\vx_j - \vx_{j-1}$. To handle these terms, we need some additional technical results. Consider the difference $\Bmat_j-\Bmat_{j-1}$. Multiplying \refeq{a-} through by $\tau_{j-1}$ gives
\begin{equation}
\begin{aligned}
\tau_{j-1} \Bmat_{j}-\tau_{j-1} \Bmat_{j-1}&=-\tau_{j-1\,s}\Gammamat_{j-1}\Bmat_{j-1} - \sum\limits_{i\neq\ell}^{j-2} \tau_{is}\Gammamat_i\Bmat_i \\
&=- \sum\limits_{i\neq\ell}^{j-1} \tau_{is}\Gammamat_i\Bmat_i.
\end{aligned}
\end{equation}
It is therefore convenient to define a new set of matrices:
\begin{equation}
\Fmat_j\equiv \tau_{j-1} \Bmat_{j}-\tau_{j-1} \Bmat_{j-1} = - \sum\limits_{i\neq\ell}^{j-1} \tau_{is}\Gammamat_i\Bmat_i
\label{eq:Fmatdef}
\end{equation}
and similarly for $\Cmat_{\ell j}$,
\begin{equation}
\Gmat_{\ell j}\equiv \tau_{j-1} \Cmat_{\ell j}-\tau_{j-1} \Cmat_{\ell\,j-1} = \tau_{\ell s}\Imat - \sum\limits_{i=\ell+1}^{j-1} \tau_{is}\Gammamat_i\Cmat_{\ell i}.
\label{eq:Gmatdef}
\end{equation}
Both the $\Fmat_j$ and $\Gmat_{\ell j}$ matrices have dimensions of time. Therefore terms in the time delay that include these matrices will not include an explicit $\tau_{ij}$ as a coefficient.

Using these relations along with the lens equation [\refeq{singlelenseq}], we find 
\begin{equation}
\tau_{j-1} \left(\vx_j - \vx_{j-1}\right) = \Fmat_j \vx_1 - \Gmat_{\ell j} \Cmat_{\ell s}^{-1} \left(\Bmat_s \vx_1 - \vx_s \right).
\label{eq:diffx}
\end{equation}
Substituting this into \refeq{t_sum}, we have
\begin{multline}
T=\frac{1}{2} \vx_\ell \cdot \left [ (\Fmat_\ell - \Fmat_{\ell+1}) \vx_1 - (\Gmat_{\ell \ell} - \Gmat_{\ell \,\ell+1}) \Cmat^{-1}_{\ell s} \left(\Bmat_s\vx_1-\vx_s \right) \right ] \\
+ \frac{1}{2}\vx_s \cdot \left[\Fmat_s \vx_1 - \Gmat_{\ell s} \Cmat^{-1}_{\ell s} (\Bmat_s \vx_1- \vx_s)\right] -\tau_{\ell s} \phi_\ell(\vx_{\ell}).
\end{multline}
Note that the $\Fmat_{j}$ do not include the main plane, so $\Fmat_\ell = \Fmat_{\ell+1}$. Also, as the $\Gmat_{\ell j}$ only include the background planes, $\Gmat_{\ell \ell}=0$ and $\Gmat_{\ell\,\ell+1} = \tau_{\ell s}$.  With these simplifications, we have our final expression:
\begin{multline}
T=\frac{1}{2}\tau_{\ell s}\Bmat_{\ell} \vx_1 \cdot \Cmat^{-1}_{\ell s} (\Bmat_s \vx_1 - \vx_s) -\tau_{\ell s} \phi_\ell(\Bmat_\ell\vx_1)\\
+ \frac{1}{2}\vx_s\cdot ( \Fmat_s \vx_1 - \Gmat_{\ell s} \Cmat^{-1}_{\ell s} \Bmat_s\vx_1  + \Gmat_{\ell s} \Cmat^{-1}_{\ell s} \vx_s) .
\label{eq:single_sch_compare}
\end{multline}
This form is most like \refeq{t_quad} and will be useful to compare to previous calculations with a single main plane. We can rewrite the result in an equivalent form that more closely resembles \refeq{t_alpha} by reordering terms and substituting for $\va_\ell$ and $\vx_\ell$:
\begin{equation}
T=\frac{1}{2}\bigl[ \vx_s \cdot \Fmat_s \vx_1 + \tau_{\ell s}\vx_\ell \cdot \va_\ell(\vx_\ell) - \vx_s \cdot \Gmat_{\ell s} \va_\ell(\vx_\ell) \bigr] - \tau_{\ell s}\phi_\ell(\vx_\ell).
\label{eq:single_final}
\end{equation}
This is the form that we will compare to our final results for multiple main planes (below).

We note that SEF previously derived an expression for the time delay in the case of a single main plane with multiple tidal planes. The analyses are complementary, because our approach is algebraic (we start with the general expression for the multiplane time delay and manipulate the expression to look for simplifications), whereas the approach in SEF is based on solving a partial differential equation. By Fermat's principle, setting the derivative of the time delay equal to zero should give the lens equation. In the case of a single main plane, the only independent variable is the position in that plane (the positions in all of the tidal planes can be written in terms of $\vx_\ell$). Therefore, the time delay must have
\begin{equation}
\frac{\partial T}{\partial \vx_\ell} \propto \Cmat^{-1}_{\ell s} \Bmat_s\Bmat^{-1}_{\ell} \vx_\ell - \Cmat^{-1}_{\ell s} \vx_s - \frac{\partial \phi_\ell}{\partial \vx_\ell}
\end{equation}
The right-hand side is linear in $\vx_\ell$, so $T$ must be quadratic, and SEF find (in our notation)
\begin{multline}
T \propto \frac{1}{2}  \Bmat_{\ell} \vx_1 \cdot \Cmat^{-1}_{\ell s} (\Bmat_s \vx_1 - \vx_s) - \frac{1}{2} \vx_{s} \cdot (\Cmat^{-1}_{\ell s})^{\Transpose} (\Bmat_\ell \vx_1 - \Bmat_\ell\Bmat^{-1}_s \vx_s)\\
- \phi_\ell(\vx_\ell) + \mbox{const}.
\label{eq:sch}
\end{multline}
Applying Fermat's principle does not specify the proportionality factor or an additive `constant' (really, any term that is independent of $\vx_\ell$). Comparing \refeqs{single_sch_compare}{sch}, we see that the proportionality constant is $\tau_{\ell s}$ \citep[which is not surprising; also see][]{Schneider97}. The terms in the first set of parentheses are identical to our solution. In the second set of parentheses, the first term in \refeq{sch} is equivalent to the corresponding term in our expression if the following identity holds:
\begin{equation}
\tau_{\ell s} (\Cmat^{-1}_{\ell s})^{\Transpose} \Bmat_\ell = \Gmat_{\ell s} \Cmat^{-1}_{\ell s} \Bmat_s -  \Fmat_s.
\label{eq:identity}
\end{equation}
The proof of this identity is given in Appendix \ref{appendix}. We note that the second term in the second set of parentheses in \refeq{sch} is not equivalent to the corresponding term in our expression, but the difference term is quadratic in $\vx_s$ and independent of $\vx_\ell$ so it is part of the `const' term in \refeq{sch}. Such a term does not affect \emph{differential} time delays, which are the observables of interest.

Thus, we conclude that our expression is equivalent to that given by SEF, at least for differential time delays. We acknowledge that our algebraic approach is more complicated than the Fermat principle argument used by SEF, at least for the case of a single main plane. However, for multiple main planes, the Fermat principle approach would require solving an arbitrarily large system of coupled partial differential equations, while our algebraic approach is easily generalized, as we are about to see. Also, our algebraic approach can pin down terms that are quadratic in $\vx_s$, which may be of formal interest even if they are unimportant for observable time delays.

\subsection{Multiple `main' planes}
\label{S:tdel-Nmain}
We now extend this analysis to an arbitrary combination of main planes and tidal planes. We again do not make any assumptions about the redshift distributions of the planes or how the planes are ordered. As above, we denote the index of main planes as $\ell \in \{\ell_1,\ell_2,\ldots,\ell_\mu,\ldots\}$. We begin with the lens equation (\ref{eq:xfinN}) for multiple main planes. We substitute this expression into the full time delay expression, \refeq{full_t}, and separate the main plane indices:
\begin{multline}
T = \sum\limits_{i=1}^{s-1} \frac{1}{2} \tau_{i} \left[\vx_{i+1} -\vx_i \right]^2
- \sum\limits_{i=1,i\not\in\{\ell_\mu\}}^{s-1} \frac{1}{2} \tau_{i} \beta_{i} \vx_i \cdot \Gammamat_i \vx_{i}\\ -\sum\limits_{\ell\in\{\ell_\mu\}}\tau_{\ell s} \phi_\ell(\vx_{\ell}).
\end{multline}
As $\Bmat_j$ and $\Cmat_{\ell j}$ only depend on the tidal planes, the relationships between these matrices and $\Gammamat_j$, \refeqs{b_g}{c_g}, still hold in the case of multiple main planes. It is useful to point out that \refeq{c_g} generalizes to each $\ell \in \{\ell_\mu\}$. Using these relations and expanding the quadratic terms, analogous to \refeq{t_sum}, we can rewrite the time delay as
\begin{multline}
T = \sum\limits_{i=1}^{s-1} \frac{1}{2} \tau_{i} \left[ \vx_{i+1}^2 - 2 \vx_{i+1}\cdot \vx_i + \vx_i^2  \right] - \sum\limits_{\ell\in\{\ell_\mu\}}\tau_{\ell s} \phi_\ell(\vx_{\ell})\\- \sum\limits_{i=1,i\not\in\{\ell_\mu\}}^{s-1} \frac{1}{2} \tau_{i} \vx_i \cdot \left[ \left(1-\frac{\tau_{i-1}}{\tau_{i}} \right ) \vx_i - \vx_{i+1} - \frac{\tau_{i-1}}{\tau_{i}}\vx_{i-1} \right].
\end{multline}
As in the single-plane case, the identity term in the second term matches the second quadratic term in the first sum. These cancel, leaving only the main planes from the first sum. Again, the $i,i+1$ cross term in the second sum cancels one of the cross terms in the first sum, leaving only the main plane terms. As in the single-plane case, we see that the remaining terms are identical but that the indices in the second sum are decremented by 1. We reindex the sums with $i\rightarrow i+1$:
\begin{equation}
\sum\limits_{i=1,i\not\in\{\ell_\mu\}}^{s-1} \tau_{i-1}  \vx_{i}^2  \rightarrow \sum\limits_{i=0,i+1\not\in\{\ell_\mu\}}^{s-2} \tau_{i}  \vx_{i+1}^2
 \end{equation}
 and
 \begin{equation}
 \sum\limits_{i=1,i\not\in\{\ell_\mu\}}^{s-1} \tau_{i-1}\vx_{i}\cdot\vx_{i-1} \rightarrow \sum\limits_{i=0,i+1\not\in\{\ell_\mu\}}^{s-2} \tau_{i}\vx_{i+1}\cdot\vx_i.
 \end{equation}
These terms now cancel in the sums leaving only the $s-1$ and the set of $\{\ell_\mu -1\}$ terms.

We are left with
 \begin{multline}
 T =  \tau_{s-1} \vx_s \cdot (\vx_s - \vx_{s-1}) + -\sum\limits_{\ell\in\{\ell_\mu\}}\tau_{\ell s} \phi_\ell(\vx_{\ell})\\
+ \frac{1}{2} \sum\limits_{\ell \in \{\ell_\mu\}} \tau_{\ell}\vx_\ell\cdot(\vx_\ell - \vx_{\ell+1}) +  \tau_{\ell-1}\vx_\ell\cdot(\vx_\ell - \vx_{\ell-1}).
\label{eq:TNtmp}
\end{multline}
The analysis proceeds as it did after \refeq{Txdiff}. Our expressions for $\Fmat_j$ and $\Gmat_{\ell j}$ remain basically unchanged except that $\ell$ becomes a free index that runs over the main planes:
 \begin{equation}
\Fmat_j\equiv \tau_{j-1} \Bmat_{j}-\tau_{j-1} \Bmat_{j-1} = - \sum\limits_{i=1,i\not \in \{\ell_\mu\}}^{j-1} \tau_{is}\Gammamat_i\Bmat_i
\label{eq:Fdef}
\end{equation}
and
\begin{equation}
\Gmat_{\ell j}\equiv \tau_{j-1} \Cmat_{\ell j}-\tau_{j-1} \Cmat_{\ell \, j-1} = \tau_{\ell s}\Imat- \sum\limits_{i=\ell+1,i \not \in \{\ell_\mu\}}^{j-1} \tau_{is}\Gammamat_i\Cmat_{\ell i}.
\label{eq:Gdef}
\end{equation}
The identity in \refeq{diffx} generalizes to 
\begin{equation}
\tau_{j-1} (\vx_j - \vx_{j-1}) = \Fmat_j \vx_1 - \sum\limits_{\ell \in \{ \ell_\mu < j\}} \Gmat_{\ell j} \va_\ell(\vx_\ell).
\end{equation}
Substituting this into \refeq{TNtmp} yields
\begin{multline}
T = \frac{1}{2} \vx_s \cdot\left[\Fmat_s \vx_1 - \sum\limits_{\ell \in \{\ell_\mu\}} \Gmat_{\ell s} \va_\ell\right] \\
+ \sum\limits_{\ell \in \{\ell_\mu\}}\left\{ \frac{1}{2} \vx_\ell \cdot \left [ \Fmat_\ell \vx_1 - \sum\limits^{\ell' < \ell}_{\ell' \in \{\ell_\mu\}} \Gmat_{\ell' \ell} \va_{\ell'} \right.\right.\\
\left.\left.+  \Fmat_{\ell+1} \vx_1 - \sum\limits^{\ell' < \ell+1}_{\ell' \in \{\ell_\mu\}} \Gmat_{\ell' \, \ell+1} \va_{\ell'}\right] - \tau_{\ell s} \phi_\ell(\vx_{\ell}) \right\}.
\end{multline}
Recall that $\Fmat_j$ and $\Gmat_{\ell j}$ are both independent of main planes so $\Fmat_\ell = \Fmat_{\ell+1}$ and $\Gmat_{\nu \ell} = \Gmat_{\nu\,\ell+1}$. Therefore, as before, the $\Fmat_j$ and $\Gmat_{\ell j}$ terms cancel. There is an important subtlety here, however. The second sum with the $\Gmat_{\nu\,\ell+1}$ includes one more main plane than the previous corresponding sum, namely $\Gmat_{\ell\,\ell+1}\va_\ell = \tau_{\ell s} \va_\ell$. We also substitute $\vx_\ell$ and $\vx_s$ from the lens \refeq{xfinN}, finally giving us
\begin{multline}
T= \frac{1}{2} \vx_s \cdot \Fmat_s \vx_1 + \sum\limits_{\ell \in \{\ell_\mu\}}\left[\frac{1}{2}\tau_{\ell s} \vx_\ell \cdot \va_\ell(\vx_\ell) \right. \\
\left.-\frac{1}{2}\vx_s \cdot \Gmat_{\ell s}\va_\ell(\vx_\ell) -\tau_{\ell s}\phi_\ell(\vx_\ell) \right].
\label{eq:TfinN}
\end{multline}
This result immediately becomes the single main plane time delay, \refeq{single_final}, by dropping the sum over main planes.

In practice, we can tabulate all of the LOS effects by calculating all of the $\Bmat_j$, $\Cmat_{\ell j}$, $\Fmat_j$, and $\Gmat_{\ell j}$ matrices. The benefit of this approach is that all of the LOS calculations can be done up front and performed only once. We can save these matrices and then vary the main plane potentials without ever having to recalculate the full LOS.

\section{Mass Sheet Degeneracy}
\label{S:masssheet}

For traditional, single-plane lensing, \citet{Falco85} showed that certain transformations of the lens potential leave the image positions and flux ratios unchanged. One notable transformation is the `mass sheet degeneracy.' In the single-plane case, the lens equation has the form
\begin{equation}
\vx_s = \vx - \nabla\phi(\vx).
\label{eq:leqn1}
\end{equation}
If we apply the transformation 
\begin{equation}
\phi(\vx) \rightarrow (1 - \kappa) \phi(\vx) + \frac{\kappa}{2} \vx^2
\label{eq:phitran}
\end{equation}
the entire right-hand side of \refeq{leqn1} gets multiplied by $(1-\kappa)$. Because the source position is unobservable, we can define a rescaled source coordinate $(1-\kappa)\vy =\vx_s$ and then write the transformed lens equation as
\begin{equation}
(1- \kappa) \vy = (1- \kappa) \vx - (1- \kappa)\nabla\phi(\vx),
\end{equation}
The $(1-\kappa)$ factors cancel, so the transformed equation is formally equivalent to the original.  A similar cancellation occurs for the fluxes if we rescale the source flux, which is permitted if the intrinsic flux of the source is unknown and constraints come from flux ratios rather than absolute fluxes.\footnote{Type Ia supernovae can be used to break the mass sheet degeneracy because their intrinsic luminosity can be inferred from their light-curve shapes \citep[e.g.][]{Kolatt98}.} Time delays are different, however. The transformation (\ref{eq:phitran}) causes differential time delays to be rescaled by
\begin{equation}
\Delta T' = (1-\kappa) \Delta T,
\end{equation}
which is important when using time delays to constrain the Hubble constant \citep[e.g.][]{Fadely10,Suyu10,Suyu13}. Overall, the mass sheet degeneracy can be viewed as a type of gauge invariance analogous to what is seen with potentials in electricity and magnetism.

Before proceeding to the multiplane case, it is useful to examine a case with external convergence and shear in the lens plane. We can write the potential as
\begin{equation}
\phi(\vx) = \phi_{\mathrm{g}}(\vx) + \frac{1}{2} \vx \cdot \Gammamat \vx
\end{equation}
where $\phi_{\mathrm{g}}(\vx)$ is the potential due to the main galaxy. The mass sheet degeneracy still applies to this situation, but the transformation is slightly different:
\begin{equation}
\phi_{\mathrm{g}}(\vx) \rightarrow (1 - \kappa) \phi(\vx) + \frac{\kappa}{2} \vx \cdot \left(\Imat - \Gammamat\right) \vx.
\label{eq:sheartrans}
\end{equation}
This form of the mass sheet degeneracy produces the same rescaling of observables as before.

We have found a similar gauge symmetry for the case of a single main plane with an arbitrary collection of tidal planes along the LOS. If we start with the lens equation (\ref{eq:xfin1}) and make the transformation
\begin{equation}
\phi(\vx_\ell) \rightarrow (1-\kappa)\phi(\vx_\ell) +\frac{\kappa}{2}\vx_\ell \cdot \Cmat^{-1}_{\ell s} \Bmat_s \Bmat_\ell^{-1} \vx_\ell
\label{eq:phi_singlemain}
\end{equation}
we find that the observables scale in the same way as the original mass sheet degeneracy. The form of this transformation is reminiscent of \refeq{sheartrans}, so we define an `effective' tidal tensor by
\begin{equation}
\Gammamat_{\rm{eff}} \equiv \Imat - \Cmat^{-1}_{\ell s} \Bmat_s \Bmat_\ell^{-1}.
\label{eq:Geff}
\end{equation}
To build some intuition about this quantity, it is useful to examine the small-shear limit. Substituting expressions from \refsec{leqn-small} yields
\begin{equation}
\Cmat^{-1}_{\ell s} \Bmat_s \Bmat_\ell^{-1} \approx ( \Imat - \Gammahat_{\rm b} )^{-1} ( \Imat - \Gammamat_{\rm tot}) (\Imat - \Gammahat_{\rm f})^{-1}.
\end{equation}
If we make the additional, stronger assumption that the sums over tidal planes are also small, we can further simplify this expression. Using a Taylor series expansion of the inverses and keeping only the first-order terms in $\Gammamat$'s, we obtain
\begin{equation}
\Cmat^{-1}_{\ell s} \Bmat_s \Bmat_\ell^{-1} \approx ( \Imat + \Gammahat_{\rm b} ) ( \Imat - \Gammamat_{\rm tot}) (\Imat + \Gammahat_{\rm f}).
\end{equation}
Multiplying this out and keeping only linear terms in $\Gammamat$'s, we find
\begin{equation}
\Gammamat_{\rm{eff}} \approx \sum\limits_{i = 1, i\neq\ell}^{N} \left(1 - \beta \right) \Gammamat_i,
\label{eq:Geff-approx}
\end{equation}
where $\beta$ is $\beta_{i \ell}$ in the foreground and $\beta_{\ell i}$ in the background. In other words, $\Gammamat_{\rm eff}$ is approximately the sum of all of the tidal planes weighted by the redshift factor $(1-\beta)$. This has the same form as the effective shear that was found by \citet{Iva}. We will comment further on the use of $\Gammamat_{\rm{eff}}$ in \refsec{conclusions}.

The mass sheet degeneracy is more subtle for multiple main planes. As an example, consider the lens equation for two main planes:
\begin{equation}
\begin{aligned}
\vx_s &= \vx_1 - \va_1(\vx_1) - \va_2(\vx_2) \\
&= \vx_1 - \va_1(\vx_1) - \va_2(\vx_1 - \beta_{1 2} \va_1(\vx_1)).
\end{aligned}
\end{equation}
Any transformation that involves a rescaling of $\va_1$ (like that in equation \ref{eq:phitran}) would create a rescaling that appears inside the argument of $\va_2$. In order for the transformation to create an overall rescaling similar to that for external convergence, the composition of the deflection functions $\va_1$ and $\va_2$ would have to be proportional to $\va_2$, which appears to be a restrictive constraint. Therefore, it remains to be seen how the multiple-main-plane mass sheet degeneracy applies in practice.

\section{Conclusions}
\label{S:conclusions}
To avoid possible biases in strong lensing studies, it is important to account for LOS effects. We have presented a lensing framework that fills the gap between using the full multiplane lens equation (which can be computationally expensive) and treating everything in the tidal approximation (which omits higher order effects that can be significant for objects that are projected near the lens and/or are massive). The framework can properly account for the non-linear effects from any mixture of `main' planes (strong lenses) that are given full treatment and `tidal' planes (weak lenses) that are treated using the tidal approximation. Our framework can be used to calculate all of the standard lensing observables. The general expressions for the lens equation, magnification tensor, and time delay are as follows (from equations \ref{eq:xfinN}, \ref{eq:AfinN}, and \ref{eq:TfinN}):
\begin{flalign*}
&\vx_{i} = \Bmat_i \vx_1 - \sum_{\ell \in \{\ell_\mu<i\}} \Cmat_{\ell i} \va_\ell(\vx_\ell), \\
&\Amat_{i} = \Bmat_i - \sum_{\ell \in \{\ell_\mu<i\}} \Cmat_{\ell i} \Gammamat_\ell \Amat_\ell, \\
&T= \frac{1}{2} \vx_s \cdot \Fmat_s \vx_1\\
&+ \sum\limits_{\ell \in \{\ell_\mu\}}\left[\frac{1}{2}\tau_{\ell s} \vx_\ell \cdot \va_\ell(\vx_\ell) -\frac{1}{2}\vx_s \cdot \Gmat_{\ell s}\va_\ell(\vx_\ell) -\tau_{\ell s}\phi_\ell(\vx_\ell) \right].
\end{flalign*}
(We emphasize that $\va_\ell$, $\Gammamat_\ell$, and $\phi_\ell$ all need to be evaluated at $\vx_\ell$, which is important for reasons discussed below.) These expressions are more accurate than what we have termed the single main plane case, because they allow for higher order effects in planes other than the main lens plane. Yet they are more efficient than the full multiplane lens equation because the recursive sums only include main planes. All of the tidal planes---which may number in the hundreds for realistic lines of sight---can be compressed into the following matrices (from equations \ref{eq:Bmatdef}, \ref{eq:Cmatdef}, \ref{eq:Fmatdef}, and \ref{eq:Gmatdef}):
\begin{align*}
\Bmat_j &= \Imat - \sum\limits_{i=1,i\not \in \{\ell_\mu\}}^{j-1}\beta_{ij}\Gammamat_i \Bmat_i, \\
\Cmat_{\ell j} &= \beta_{\ell j} \Imat - \sum\limits_{i=\ell+1, i\not\in\{\ell_\mu\}}^{j-1} \beta_{i j}\Gammamat_i\Cmat_{\ell i}, \\
\Fmat_j &\equiv \tau_{j-1} \Bmat_{j}-\tau_{j-1} \Bmat_{j-1}
  \ =\ - \sum\limits_{i=1,i\not \in \{\ell_\mu\}}^{j-1} \tau_{is}\Gammamat_i\Bmat_i, \\
\Gmat_{\ell j}&\equiv \tau_{j-1} \Cmat_{\ell j}-\tau_{j-1} \Cmat_{\ell \,j-1}
  \ =\ \tau_{\ell s}\Imat- \sum\limits_{i=\ell+1,i \not \in \{\ell_\mu\}}^{j-1} \tau_{is}\Gammamat_i\Cmat_{\ell i}.
\end{align*}
These matrices can be computed once at the start of any lensing analysis and stored for repeated use.

To date, a common modelling approach has been to incorporate the main lens galaxy and any strong perturbers (assumed to lie in the same plane as the lens) into $\va_\ell$, to fit for an external shear in the main plane, and then to correct for remaining LOS effects through an external convergence (e.g. \citealt{Hilbert09, Suyu10, Collett13, Suyu13}, but see \citealt{Schneider13}). Our analysis leads to two remarks. First, these LOS convergence corrections are been calibrated by ray tracing through cosmological simulations to compute the total convergence from a direct sum of all the mass along the LOS. We find, however, that the key quantities are the effective convergence and shear, which are given by (from equations \ref{eq:Geff} and \ref{eq:Geff-approx})
\begin{eqnarray*}
\Gammamat_{\rm{eff}} \ \equiv\ \Imat - \Cmat^{-1}_{\ell s} \Bmat_s \Bmat_\ell^{-1}
\ \approx\ \sum\limits_{i = 1, i\neq\ell}^{N} (1 - \beta) \Gammamat_i ,
\end{eqnarray*}
where $\beta$ is $\beta_{i \ell}$ in the foreground of the main lens plane, and $\beta_{\ell i}$ in the background. The $\beta$ weighting factors depend on the redshift of the main lens galaxy as well as the redshifts of the source and the plane in question, so the effective shear and convergence cannot be tabulated in a general way that is independent of particular lens systems.

Secondly, most existing lens models have been fit to the positions of the images on the sky (which we have denoted by $\vx_1$). Each main plane actually needs to be evaluated using the position $\vx_\ell$ of the light ray in that plane. This distinction gives rise to non-linearities that cannot be mimicked by a simple shear and can lead to systematic uncertainties in lens models if not handled properly (McCully et al., in preparation). In principle, the `corrective' approach to lens modelling could account for the non-linear effects by using $\vx_\ell = \Bmat_\ell \vx_1$, where the matrix $\Bmat_\ell$ can be calibrated by ray tracing.

A different approach to lens modelling is to directly incorporate LOS effects by building full three-dimensional mass models like those used by \citet{Wong}. Then all of the non-linear effects are automatically included, and the convergence and shear are computed self-consistently from an underlying mass distribution. In order to employ our hybrid framework effectively, we need to understand when it is acceptable to use the tidal approximation and when we need to treat a plane exactly. In a forthcoming paper (McCully et al., in preparation), we use realistic beams like those in \citet{Wong} to test the tidal approximation. We also quantify bias and scatter in lens models associated with different ways of handling the LOS. The framework presented here serves as the foundation for such detailed treatments of LOS effects in strong lensing.

\section*{Acknowledgements}
We thank the referee, Peter Schneider, for very detailed and helpful comments. We thank Phil Marshall, Roger Blandford, Sherry Suyu, and Stefan Hilbert for helpful conversations.
CM and CRK acknowledge funding from NSF grants AST-0747311 and AST-1211385.
KCW is supported by an EACOA Fellowship awarded by the East Asia Core Observatories Association, which consists of the Academia Sinica Institute of Astronomy and Astrophysics, the National Astronomical Observatory of Japan, the National Astronomical Observatory of China, and the Korea Astronomy and Space Science Institute.
AIZ acknowledges funding from NSF grants AST-0908280 and AST-1211874, as well as NASA grants
ADP-NNX10AD476 and ADP-NNX10AE88G.
She also thanks the John Simon Guggenheim Memorial Foundation and the Center for Cosmology and Particle Physics at NYU for their support.
Image credits: Centaurus A---Jean-Charles Cuillandre, Giovanni Anselmi, Hawaiian Starlight; Leo I---Oliver Stein.


\bibliographystyle{mn2e}
\bibliography{multiplane-full}

\appendix
\section{Matrix Identities}\label{appendix}

In Section \ref{S:tdel-1main}, we found that our expression for the time delay is equivalent to an alternative expression found by SEF only if the following identity holds (see equation \ref{eq:identity}):

\begin{equation}
\tau_{\ell s} (\Cmat^{-1}_{\ell s})^\Transpose \Bmat_\ell = \Gmat_{\ell s} \Cmat^{-1}_{\ell s} \Bmat_s -  \Fmat_s.
\end{equation}
Moving $\Cmat^{-1}_{\ell s}$ to the right-hand side and using the definitions of $\Fmat_s$ and $\Gmat_{\ell s}$ from \refeqs{Fdef}{Gdef} yields
\begin{equation}
\begin{aligned}
\tau_{\ell s} \Bmat_\ell
&= \tau_{s-1} \Cmat_{\ell s}^\Transpose \left[ \left(\Cmat_{\ell s} - \Cmat_{\ell \, s-1} \right ) \Cmat^{-1}_{\ell s} \Bmat_s  - \left(\Bmat_s - \Bmat_{s-1}\right)\right] \\
&= \tau_{s-1} \Cmat_{\ell s}^\Transpose \left( \Bmat_{s-1} - \Cmat_{\ell \, s -1} \Cmat^{-1}_{\ell s} \Bmat_s \right).
\end{aligned}
\label{eq:identity2}
\end{equation}
This is the form of the identity we seek to prove.

We find it helpful to begin with two special cases. First, if all tidal planes are in the foreground then $\ell = s-1$ so $\tau_{s-1} = \tau_{\ell s}$ and $\Bmat_{s-1} = \Bmat_{\ell}$. Also, the background matrices are $\Cmat_{\ell s} = \Imat$ and $\Cmat_{\ell \, s-1} = \zeromat$. Therefore the right-hand side of \refeq{identity2} reduces to $\tau_{\ell s} \Bmat_{\ell}$, which proves the identity for this case.  Second, consider a single tidal plane that lies in the background and is characterized by the tidal matrix $\Gammamat$. In this case, the matrices are as follows:
\begin{gather}
\Bmat_\ell = \Imat , 
\qquad
\Bmat_{s-1} = \Imat ,
\qquad
\Bmat_s = \Imat - \Gammamat ,\nonumber
\\
\Cmat_{\ell \, s-1} = \beta \Imat ,
\qquad
\rm{and} \qquad \Cmat_{\ell s} = \Imat - \beta \Gammamat.
\end{gather}
Therefore the left-hand side of \refeq{identity2} is $\tau_{\ell s} \Imat$, while the right-hand side is
\begin{eqnarray}
\mbox{RHS}
&=& \tau_{s-1} (\Imat-\beta\Gammamat) \left[ \Imat - \beta (\Imat-\beta\Gammamat)^{-1} (\Imat-\Gammamat) \right] \nonumber \\
&=& \tau_{s-1} \left[ (\Imat-\beta\Gammamat) - \beta (\Imat-\Gammamat) \right] \nonumber \\
&=& \tau_{s-1} (1-\beta) \Imat .
\end{eqnarray}
Using \refeqset{b_1}{t_ik} we find $\tau_{s-1} (1-\beta) = \tau_{\ell s}$, which proves the identity for this case.

To prove the identity (\ref{eq:identity2}) in general, we need to review some ancillary results and establish some new ones.  From \refeq{Bjidentity} and the ensuing discussion, we have recursion relations for $\Bmat_j$ and $\Cmat_{\ell j}$:
\begin{eqnarray}
\Bmat_{j+1} &=& \Mmat_j \Bmat_j - \frac{\tau_{j-1}}{\tau_{j}} \Bmat_{j-1} , \label{eq:Brecur}\\
\Cmat_{\ell \, j+1} &=& \Mmat_j \Cmat_{\ell \, j} - \frac{\tau_{j-1}}{\tau_{j}} \Cmat_{\ell \, j-1} ,  \label{eq:Crecur}
\end{eqnarray}
where we define
\begin{equation}
\Mmat_j \equiv \left(1+\frac{\tau_{j-1}}{\tau_{j}}\right) \Imat - \beta_j \Gammamat_j .
\label{eq:Mmatdef}
\end{equation}
Note that $\Gammamat_j$ is symmetric and so $\Mmat_j$ is symmetric as well.  We define a new matrix with the structure that we are looking for on the right-hand side of \refeq{identity2}:
\begin{equation}
\Wmat_j \equiv \Bmat_j - \Cmat_{\ell j} \Cmat_{\ell s}^{-1} \Bmat_s .
\label{eq:Wmatdef}
\end{equation}
We can combine \refeqs{Brecur}{Crecur} to write a recursion relation for $\Wmat_j$:
\begin{equation}
\Wmat_{j+1} = \Mmat_j \Wmat_j - \frac{\tau_{j-1}}{\tau_{j}} \Wmat_{j-1} .
\label{eq:Wrecur}
\end{equation}

It is convenient to define a generalized version of the $\Cmat$ matrices (compare equation~\ref{eq:Cmatdef}),
\begin{equation}
\Cmat_{ik} \equiv \beta_{ik} \Imat - \sum_{j=i+1}^{k-1} \beta_{jk} \Gammamat_j \Cmat_{ij} .
\end{equation}
Note that the sum runs over the second index of $\Cmat$. Because $\Cmat_{ij}$ contains $\Gammamat$ matrices between planes $i$ and $j$, the products of $\Gammamat$ matrices have indices that decrease to the right.  If we restrict attention to matrices $\Cmat_{is}$ in which the second index is $s$, we can write an alternative form of the sum as
\begin{equation}
\begin{aligned}
\Cmat_{is} \ &=\ \Imat - \sum_{j=i+1}^{s-1} \beta_{ij} \Cmat_{js} \Gammamat_j\\
&=\ \Imat + \sum_{j=i+1}^{s-1} \frac{\beta_{ij}}{\beta_{j}} \Cmat_{js} \left[ \Mmat_j - \left(1+\frac{\tau_{j-1}}{\tau_{j}}\right) \Imat \right] .
\end{aligned}
\end{equation}
Here, the sum runs over the first index of $\Cmat$, and because $\Cmat_{js}$ contains $\Gammamat$ matrices between planes $j$ and $s$, the factor of $\Gammamat_j$ needs to be on the right in order to have the matrix product arranged with indices that decrease to the right. In the second step, we use \refeq{Mmatdef} to replace $\Gammamat_j$ with $\Mmat_j$. Note that
\begin{multline}
\Cmat_{i-1 \, s} - \Cmat_{is}
= \frac{\beta_{i-1}}{\beta_{i}} \Cmat_{is} \left[ \Mmat_{i} - \left(1+\frac{\tau_{i-1}}{\tau_{i}}\right) \Imat \right]\\
+ (\tau_{is} - \tau_{i-1 \, s}) \sum_{j=i+1}^{s-1} \frac{1}{\beta_{j} \tau_{js}} \Cmat_{js} \left[ \Mmat_{j} - \left(1+\frac{\tau_{j-1}}{\tau_{j}}\right) \Imat \right] ,
\label{eq:Ctmp}
\end{multline}
where we make use of \refeqset{b_1}{t_ik}.  We can write a similar expression for $\Cmat_{is} - \Cmat_{i+1 \, s}$ and combine it with \refeq{Ctmp} to eliminate the sum. Again using \refeqset{b_1}{t_ik} to simplify yields
\begin{equation}
\Cmat_{i-1 \, s} = \frac{\beta_{i-1}}{\beta_{i}} \left( \Cmat_{is} \Mmat_i - \frac{\tau_{is}}{\tau_{i+1 \, s}} \Cmat_{i+1 \, s} \right).
\label{eq:Crecur2}
\end{equation}
We can simplify one step further by introducing a scaled version of the matrices:
\begin{equation}
\Chat_i \equiv \frac{1}{\beta_i} \Cmat_{is} .
\label{eq:Chatdef}
\end{equation}
With this definition, \refeq{Crecur2} becomes
\begin{equation}
\Chat_{i-1} = \Chat_i \Mmat_i - \frac{\tau_{i}}{\tau_{i+1}} \Chat_{i+1},
\label{eq:Crecur3}
\end{equation}
where we use \refeq{t_b} to put $\tau_{is}/\beta_i = \tau_i$ (and similar for index $i+1$). If we start from index $s$ and work our way down, the first few matrices are
\begin{align}
\Chat_{s-1} &= \Imat, \label{eq:Cs-1}\\
\Chat_{s-2} &= \Mmat_{s-1}, \label{eq:Cs-2}\\
\Chat_{s-3} &=\Mmat_{s-1} \Mmat_{s-2} - \frac{\tau_{s-2}}{\tau_{s-1}} \Imat, \label{eq:Cs-3}\\
\Chat_{s-4} &= \Mmat_{s-1} \Mmat_{s-2} \Mmat_{s-3} - \frac{\tau_{s-2}}{\tau_{s-1}} \Mmat_{s-3} - \frac{\tau_{s-3}}{\tau_{s-2}} \Mmat_{s-1}. \label{eq:Cs-4}
\end{align}

There is one more useful technical result:
\begin{equation}
\Chat_{j+1} \Chat_j^{\Transpose} = \Chat_j \Chat_{j+1}^{\Transpose}.
\label{eq:Csymm}
\end{equation}
We prove this by induction (see \citealt{Seitz94} for a similar argument).  The relation is trivial for $j=s-2$ because $\Chat_{s-1} = \Imat$. It is manifestly true for $j=s-3$ because we can evaluate the left- and right-hand sides explicitly using \refeqs{Cs-2}{Cs-3}:
\begin{align}
\mbox{LHS} &= \Mmat_{s-1} \left( \Mmat_{s-2} \Mmat_{s-1} - \frac{\tau_{s-2}}{\tau_{s-1}} \Imat \right) \nonumber\\
&=\ \Mmat_{s-1} \Mmat_{s-2} \Mmat_{s-1} - \frac{\tau_{s-2}}{\tau_{s-1}} \Mmat_{s-1}, \\
\mbox{RHS} &= \left( \Mmat_{s-1} \Mmat_{s-2} - \frac{\tau_{s-2}}{\tau_{s-1}} \Imat \right) \Mmat_{s-1}\nonumber\\
 &= \Mmat_{s-1} \Mmat_{s-2} \Mmat_{s-1} - \frac{\tau_{s-2}}{\tau_{s-1}} \Mmat_{s-1}.
\end{align}
(Note that because $\Mmat_j$ is symmetric, $(\Mmat_{s-1} \Mmat_{s-2})^{\Transpose} = \Mmat_{s-2} \Mmat_{s-1}$.)  Now, if we postulate that \refeq{Csymm} is true for index $j$, we can ask what it implies for index $j-1$:
\begin{eqnarray}
\Chat_j \Chat_{j-1}^{\Transpose}
&=& \Chat_j \left( \Mmat_j \Chat_j^{\Transpose} - \frac{\tau_{j}}{\tau_{j+1}} \Chat_{j+1}^{\Transpose} \right) \nonumber\\
&=& \left( \Chat_j \Mmat_j - \frac{\tau_{j}}{\tau_{j+1}} \Chat_{j+1} \right) \Chat_j^{\Transpose} \nonumber \\
&=& \Chat_{j-1} \Chat_j^{\Transpose}.
\end{eqnarray}
In the first line we use \refeq{Crecur3} for $\Chat_{j-1}^{\Transpose}$, and in the second line we use \refeq{Csymm} to replace $\Chat_j \Chat_{j+1}^{\Transpose}$ with $\Chat_{j+1} \Chat_j^{\Transpose}$ according to our postulate.  We see that if \refeq{Csymm} is true for index $j$ then it is also true for index $j-1$, which completes the proof by induction.

Now we have all the pieces needed to prove \refeq{identity2}.  We start with a trivial relation from \refeq{Wmatdef} with $j=s$:
\begin{equation}
\zeromat = \Wmat_s .
\end{equation}
We use \refeq{Wrecur} on the right-hand side:
\begin{equation}
\zeromat = \Mmat_{s-1} \Wmat_{s-1} - \frac{\tau_{s-2}}{\tau_{s-1}} \Wmat_{s-2} .
\label{eq:Wtmp1}
\end{equation}
We can solve this to find
\begin{equation}
\Wmat_{s-2} = \frac{\tau_{s-1}}{\tau_{s-2}} \Chat_{s-2}^{\Transpose} \Wmat_{s-1} ,
\label{eq:Ws-2}
\end{equation}
where we use \refeq{Cs-2} to write this in a form involving $\Chat$, for reasons that will become clear.  Now, we return to \refeq{Wtmp1}, again apply \refeq{Wrecur} to the first term, and rearrange to find
\begin{equation}
\begin{aligned}
\zeromat  &=\ \left( \Mmat_{s-1} \Mmat_{s-2} - \frac{\tau_{s-2}}{\tau_{s-1}} \Imat \right) \Wmat_{s-2} - \frac{\tau_{s-3}}{\tau_{s-2}} \Mmat_{s-1} \Wmat_{s-3}\\
 &=\ \Chat_{s-3} \Wmat_{s-2} - \frac{\tau_{s-3}}{\tau_{s-2}} \Chat_{s-2} \Wmat_{s-3} , 
 \end{aligned}
\end{equation}
where we use \refeqs{Cs-2}{Cs-3}.  We solve for $\Wmat_{s-3}$ and use \refeq{Ws-2}:
\begin{align}
\Wmat_{s-3} &= \frac{\tau_{s-1}}{\tau_{s-3}} \Chat_{s-2}^{-1} \Chat_{s-3} \Chat_{s-2}^{\Transpose} \Wmat_{s-1} \nonumber \\
&= \frac{\tau_{s-1}}{\tau_{s-3}} \Chat_{s-2}^{-1} \left( \Chat_{s-2} \Mmat_{s-2} - \frac{\tau_{s-2}}{\tau_{s-1}} \Chat_{s-1} \right) \Chat_{s-2}^{\Transpose} \Wmat_{s-1} \nonumber \\
&= \frac{\tau_{s-1}}{\tau_{s-3}} \left( \Mmat_{s-2} \Chat_{s-2}^{\Transpose} - \frac{\tau_{s-2}}{\tau_{s-1}} \Chat_{s-1}^{\Transpose} \right) \Wmat_{s-1} \nonumber\\
&= \frac{\tau_{s-1}}{\tau_{s-3}} \Chat_{s-3}^{\Transpose} \Wmat_{s-1}.
\label{eq:Ws-3}
\end{align}
We use \refeq{Crecur3} in the second step, \refeq{Csymm} in the third step, and \refeq{Crecur3} again in the fourth step.  Repeating the analysis reveals the pattern that $\Wmat_j$ can be written as
\begin{equation}
\Wmat_j = \frac{\tau_{s-1}}{\tau_{j}} \Chat_j^{\Transpose} \Wmat_{s-1} .
\label{eq:Wj}
\end{equation}
We can prove this result by induction.  First, repeatedly applying \refeq{Crecur3} to \refeq{Wtmp1} yields
\begin{equation}
\zeromat = \Chat_{j-1} \Wmat_j - \frac{\tau_{j-1}}{\tau_{j}} \Chat_j \Wmat_{j-1} ,
\end{equation}
which we can solve to find
\begin{equation}
\Wmat_{j-1} = \frac{\tau_{j}}{\tau_{j-1}} \Chat_j^{-1} \Chat_{j-1} \Wmat_j .
\end{equation}
If we postulate that \refeq{Wj} holds for index $j$, we can write
\begin{eqnarray}
\Wmat_{j-1} &=& \frac{\tau_{s-1}}{\tau_{j-1}} \Chat_j^{-1} \Chat_{j-1} \Chat_j^{\Transpose} \Wmat_{s-1} \nonumber\\
&=& \frac{\tau_{s-1}}{\tau_{j-1}} \Chat_j^{-1} \left( \Chat_j \Mmat_j - \frac{\tau_{j}}{\tau_{j+1}} \Chat_{j+1} \right) \Chat_j^{\Transpose} \Wmat_{s-1} \nonumber\\
&=& \frac{\tau_{s-1}}{\tau_{j-1}}  \left( \Mmat_j \Chat_j^{\Transpose} - \frac{\tau_{j}}{\tau_{j+1}} \Chat_{j+1}^{\Transpose} \right) \Wmat_{s-1} \nonumber \\
&=& \frac{\tau_{s-1}}{\tau_{j-1}} \Chat_{j-1}^{\Transpose} \Wmat_{s-1}.
\end{eqnarray}
Therefore if \refeq{Wj} holds for index $j$ then it also holds for index $j-1$, which completes the proof by induction.

To finish the full proof, we use \refeq{Chatdef} to write
\begin{equation}
\frac{1}{\tau_j} \Chat_j^{\Transpose} = \frac{1}{\beta_j \tau_j} \Cmat_{js}^{\Transpose} = \frac{1}{\tau_{js}} \Cmat_{js}^{\Transpose} ,
\end{equation}
using \refeq{t_b}.  Then \refeq{Wj} becomes
\begin{equation}
\tau_{js} \Wmat_j = \tau_{s-1} \Cmat_{js}^{\Transpose} \Wmat_{s-1} .
\end{equation}
We evaluate this at $j = \ell$ and use $\Wmat_\ell = \Bmat_\ell$, which holds because $\Cmat_{\ell\ell} = \zeromat$.  This yields our final result
\begin{equation}
\tau_{\ell s} \Bmat_\ell = \tau_{s-1} \Cmat_{\ell s}^{\Transpose} (\Bmat_{s-1} - \Cmat_{\ell \, s-1} \Cmat_{\ell s}^{-1} \Bmat_s ) ,
\end{equation}
which is the identity we sought to prove.

\end{document}